# Temporal delivery, biological context and predictive modelling of normal tissue sparing in FLASH radiotherapy

Lucy Whitmore[1], Angela Corvino[1], Adel Zaid I. Mutahar[1], Banita Thakur[1], Xianjin Dai[2], Billy W. Loo Jr[2], Frederick M. Dirbas[1]

[1] Department of Surgery, Stanford University School of Medicine, Stanford Cancer Institute, Stanford, CA 94305, USA

[2] Department of Radiation Oncology, Stanford University School of Medicine, Stanford, CA 94305, USA

Corresponding author: Frederick M. Dirbas. Co-principal investigators: Billy W. Loo Jr and Frederick M. Dirbas.

## Abstract

Objective. To investigate the physical and biological factors associated with normal tissue sparing (NTS) in FLASH radiotherapy, assess the predictive value of delivery parameters beyond dose and tissue type, and evaluate the generalisability of candidate delivery thresholds.

Approach. We identified 101 preclinical in vivo publications comparing FLASH and conventional irradiation through October 2026. NTS was assessed across tissues, doses and treatment schedules. Dose-modifying factors (DMFs) and adverse event rates quantified sparing and residual injury. Regularised logistic regression with leave-one-publication-out validation assessed electron delivery parameters. Candidate thresholds were evaluated using publication-grouped five-fold cross-validation. Sensitivity analyses examined the influence of outcome definition and documented conventional tissue injury.

Main results. NTS was reported in 254/383 irradiation conditions (66.3%) across 79 publications, with higher frequencies following single exposures (68.5%) than multi-day treatments (42.4%). Publication-balanced mean DMFs for single exposures were 1.38 for skin and 1.09 for gastrointestinal injury. Adding dose per pulse (DPP) to models containing dose per fraction and tissue type increased the area under the receiver operating characteristic curve (AUROC) from 0.421 to 0.583 for single exposures and from 0.511 to 0.683 when multi-day treatments were included. Restricting analyses to conditions with documented conventional injury reduced the corresponding AUROC improvements from 0.162 to 0.086 and from 0.172 to 0.098, with both 95% confidence intervals including zero. Candidate DPP thresholds varied with schedule inclusion and showed limited discrimination across publications.

Significance. DPP improved prediction of reported NTS beyond dose and tissue type, highlighting the relevance of temporal beam structure in electron FLASH radiotherapy. These findings provide a quantitative basis for future experimental design and the refinement of FLASH delivery criteria that account for biological response and treatment schedule.

## 1. Introduction

FLASH radiotherapy delivers radiation at ultrahigh dose rates and has the potential to reduce normal tissue toxicity while maintaining tumour control, as demonstrated across a wide range of preclinical models. Normal tissue sparing (NTS) has been reported in skin, gastrointestinal (GI) tissue, lung, brain and heart with electron, proton, photon and heavy-ion beams (Favaudon et al 2014; Montay-Gruel et al 2017; Simmons et al 2019; Ruan et al 2021; Velalopoulou et al 2021; Tinganelli et al 2022; Shi et al 2022; Sørensen et al 2022a, 2022b; Kim et al 2024; Kristensen et al 2025a). However, NTS is not consistently observed across experiments using ultrahigh dose rates. Establishing the conditions under which NTS can be reliably reproduced across experimental settings is essential to guide the clinical translation of FLASH radiotherapy.

Average dose rate is widely used to describe FLASH irradiation but does not capture its temporal structure. In pulsed electron beams, the same average rate can arise from different doses per pulse (DPP), pulse durations, repetition frequencies and pulse counts. Scanned proton beams add spatial variation, complicating comparisons between field-average and local exposure rates (Esplen et al 2020; Wilson et al 2020). Accordingly, studies have identified different temporal parameters as influential under different irradiation conditions (Liu et al 2025a, 2025b; Grilj et al 2025; Zhu et al 2025).

Biological context also matters. Total dose, dose per fraction, irradiated volume, animal model and follow-up affect both conventional injury and the probability of detecting NTS. Preserved function, reduced histological damage and lower treatment-related mortality are distinct outcomes; benefit in one does not imply benefit in another. Binary NTS classifications therefore need to be considered alongside quantitative injury measures and isoeffect dose ratios (Böhlen et al 2022; Limoli and Vozenin 2023). At an individual dose, minimal or saturated injury can obscure a benefit despite a shift between the underlying dose–response curves.

Treatment schedules distribute dose over different timescales, from single exposures and intra-day splits to multi-day and mixed FLASH/conventional courses. Skin and intestinal dose-response studies suggest that NTS can diminish with fractionation, whereas functional benefits have been reported after fractionated brain irradiation (Limoli et al 2023; Ilina et al 2025; Kristensen et al 2025c; Sesink et al 2026a). Fraction number must therefore be considered alongside fraction size, timing and endpoint.

Hecker et al (2026) benchmarked isoeffective dose ratios and temporal-profile models using electron and proton data, principally acute murine skin toxicity, with intestinal and zebrafish data for comparison. Colizzi et al (2026) reviewed 50 murine electron and proton publications, comparing survival, crypt regeneration, moist desquamation and novel-object recognition. Their endpoint-specific regression analyses used publication-group-held-out validation and bootstrapping and showed substantial group sensitivity. These studies motivate evaluating delivery information alongside biological heterogeneity and the transferability of predictive rules.

We mapped the occurrence and magnitude of NTS, examined variation with treatment schedule and experimental context, and tested whether electron delivery parameters added predictive information beyond dose and tissue in held-out publications. We also assessed sensitivity to outcome definition and documented conventional injury, alongside the reproducibility of candidate thresholds across heterogeneous experimental settings.

## 2. Methods

### 2.1. Study identification and eligibility

We identified studies published between January 2014 and October 2026 using PubMed, Scopus, Google Scholar and Europe PMC. Search terms and available search records are described in supplementary section S1.1. Eligible studies were full-text publications or preprints reporting nonhuman in vivo irradiation with an interpretable comparator at conventional dose rate, including comparisons supported by a conventional dose–response curve. We excluded human studies, insects, zebrafish embryos, experiments confined to in vitro models and abstracts without full reports.

Preprints and subsequent publications were linked, and duplicate observations were removed. Publications reporting shared experimental cohorts were assigned to the same validation group so that related observations remained together during subsequent analyses.

### 2.2. Data extraction and analytical units

For each eligible comparison, we recorded tissue, animal model, sex, interventions, irradiation geometry, treatment schedule, endpoint and assessment time. Delivery variables included total dose, dose per fraction, DPP, average and intrapulse dose rate, pulse width, pulse count, repetition frequency and irradiation time. Reported values were distinguished from calculated values and qualified protocol estimates (supplementary section S4 and table S3).

Treatment schedules were classified according to how dose was distributed over time. A single exposure comprised one session without deliberate dose splitting. Homogeneous multi-day schedules used the same irradiation modality across treatment days, including courses with twice-daily fractions. Intra-day splitting, mixed FLASH/conventional courses and reirradiation were retained as separate categories (supplementary table S1).

An endpoint comparison represented a normal tissue outcome at a specified dose, irradiation setting, tissue and assessment time. For delivery analyses, compatible endpoints were grouped into irradiation conditions using dose, schedule, beam delivery, field geometry, animal model, sex and intervention. Each tested dose was a separate condition when its outcome could be adjudicated; independently reported sexes also remained separate. Assays, assessment times or animal cohorts could contribute to the same condition when the irradiation protocol was documented as common.

Conditions were also grouped within tissue and assessment time strata. These could overlap and need not sum to the overall condition count. Analytical units and cohorts for analyses of delivery alone are described in supplementary section S2 and figure S1.

### 2.3. Classification of normal tissue sparing

We classified NTS from the source authors' reported interpretation of each evaluable normal tissue endpoint. Descriptive benefits retained their statistical qualifications. A favourable point estimate alone was insufficient for a positive classification. Uninvestigated or uninterpretable endpoints were excluded from the binary outcome. Tumour and mechanistic findings remained in the evidence map but did not determine the normal tissue classification.

Under the primary definition, an irradiation condition was positive if NTS was reported in at least one evaluable endpoint. Conditions containing both positive and non-positive endpoints were flagged as discordant but remained positive under this any-NTS rule. A condition was negative when every evaluable endpoint was unchanged or worse. This classification did not establish equivalence between FLASH and conventional irradiation. It describes reported benefit at the tested dose and assessed endpoints, rather than whether the underlying FLASH and conventional dose–response curves differ.

To assess dependence on the outcome definition, we generated two alternative classifications: one required NTS in every evaluable endpoint, and the other excluded discordant conditions. The same definitions were applied to recorded discordance in linked dose–response comparisons (supplementary section S2.1).

### 2.4. Quantification of NTS magnitude

The magnitude of NTS was assessed using dose-modifying factors (DMFs), matched adverse event risks and paired functional measurements. DMF was defined as the ratio of FLASH to conventional (CONV) dose required to produce equivalent injury for the same endpoint and assessment time:

$$DMF = \frac{D_{FLASH}}{D_{CONV}} \quad (at\ equivalent\ injury)$$

A DMF greater than one indicated that a higher FLASH dose was required to produce the same level of injury. Where publications reported the reciprocal ratio, estimates and corresponding interval limits were inverted. Each distinct DMF was included once. For each tissue, estimates were first averaged within publications or shared-cohort groups, with equal weighting subsequently applied across groups. Treatment schedule was classified according to the underlying dose–response experiment (supplementary section S9 and table S9).

DMFs were obtained from published dose–response fits, reported isoeffect estimates or qualified matched-effect comparisons. Only studies providing an interpretable isoeffect estimate were included, and dose points from separate publications were not combined to construct dose–response curves. The injury level at which DMFs were evaluated varied between studies and was not necessarily equivalent to 50% normal tissue complication probability (NTCP50).

The availability of paired FLASH and CONV dose–response data was assessed across the comparative publications using extracted outcomes and targeted examination of the original studies. We identified publications reporting observed responses below and above 50% in both modalities and determined which also reported a fitted dose–response curve with at least three dose groups per modality. These criteria characterised dose–response coverage rather than the precision of individual NTCP50 estimates (supplementary section S9.3).

Adverse event analyses included matched FLASH and CONV comparisons reporting a defined harmful endpoint, event counts and animal denominators. Event risk and absolute risk reduction (ARR) were calculated as:

$$Event\ risk(\%) = \left(\frac{number\ of\ animals\ with\ event}{number\ of\ evaluable\ animals}\right) \times 100$$

$$ARR(percentage\ points) = Risk_{CONV}(\%) - Risk_{FLASH}\,(\%)$$

Absolute event risk following FLASH was also reported to quantify residual injury. Mortality and toxicity-related euthanasia were analysed separately from nonfatal events. Comparisons in which neither group or all animals in both groups experienced the event were retained. Confidence intervals for risk differences were calculated using the independent-binomial Newcombe method (Newcombe 1998). Risk ratio confidence intervals were calculated on the logarithmic scale when events occurred in both groups.

Comparisons were presented by tissue and treatment schedule without pooling animals across studies or estimating a common treatment effect. Follow-up duration, denominator restrictions, shared controls and reconstructed event counts are detailed in supplementary section S14.

Functional outcomes were assessed using paired quantitative measurements. Differences were calculated on the original measurement scale where possible; otherwise, ratios reported in the original publications were retained. Comparisons were grouped by measurement type and assessment time, distinguishing between means, medians, proportions and endpoint-free estimates. Repeated measurements were linked to their respective experimental cohorts, and different functional measures were analysed separately (supplementary section S14). Ordinal injury outcomes were classified according to the criteria described in supplementary section S3.

### 2.5. Descriptive and association analyses

The frequency of NTS was calculated at the endpoint and irradiation-condition levels and summarised by tissue and treatment schedule. Outcomes were grouped according to assessment time (≤7, >7–30, >30–90 or >90 days). Outcomes spanning multiple time windows, survival and event follow-up data, and outcomes with uncertain assessment times were considered separately.

For analyses of assessment timing and inspired gas, conditions were pooled within each category. Irradiated field areas were summarised separately for each tissue. Descriptive 95% confidence intervals were calculated using 2,000 bootstrap resamples at the publication or shared-cohort level to preserve dependencies between observations. Confidence intervals were not calculated for categories containing fewer than five groups or lacking informative between-group variation.

For single-exposure experiments, the minimum and maximum doses associated with and without NTS were determined for each tissue. This analysis was restricted to conditions with an established target-organ dose and outcomes attributable to that organ. Assessment timing and dose eligibility criteria are detailed in supplementary sections S12–S13 and table S11.

Associations between NTS and physical or experimental parameters were examined using logistic generalised estimating equations with an independence working correlation structure and uncertainty clustered by publication or shared-cohort group (Liang and Zeger 1986). Each group was assigned equal total weight, distributed equally across its irradiation conditions and, where applicable, across endpoints within each condition.

Models incorporating all beam types were adjusted for tissue, beam type and treatment schedule. Dose rate models were additionally adjusted for dose per fraction. Associations with physical parameters were expressed per doubling of the predictor. Multiple comparisons were accounted for using the Benjamini–Hochberg procedure within each family of exploratory analyses (Benjamini and Hochberg 1995). Model eligibility criteria and the treatment of sparsely represented categories are described in supplementary section S5.

### 2.6. Machine-learning prediction and cutoff validation

Predictive models were used to determine whether dose per pulse (DPP) and average dose rate predicted NTS individually or in combination, and whether they improved prediction beyond dose per fraction and tissue type (table 1). Electron irradiation experiments were analysed in two cohorts: single exposures alone and single exposures combined with homogeneous multi-day treatment schedules. Experiments investigating a range of dose rates were included without imposing a minimum dose rate, while conventional reference-only arms were excluded.

Models containing delivery parameters alone included comparisons for which the relevant parameters were available. Dose–response experiments with constant delivery parameters contributed one comparison per curve, rather than separate observations at each dose. Direct comparisons between models were performed using identical sets of observations. Models incorporating dose per fraction and tissue type required a unique, positive dose per fraction and values for both delivery parameters. Curve-level comparisons and one irradiation condition without a unique dose were therefore excluded.

L2-regularised logistic regression ($C = 1$) was used to assess the incremental predictive value of delivery parameters. Given the number of independent publication groups available for these comparisons (32–36), a fixed model specification was used without hyperparameter tuning. Regularisation limited the magnitude of fitted coefficients.

Positive continuous predictors were $\log_{10}$-transformed and standardised using the training data within each validation fold. Categorical indicators for tissue and treatment schedule were left unscaled.

Leave-one-publication-out (LOPO) cross-validation was performed by withholding each publication or shared-cohort group in turn and fitting the model to the remaining groups. Discrimination was assessed using the area under the receiver operating characteristic curve (AUROC), calculated from the combined held-out predictions. Sensitivity, specificity and balanced accuracy (BA), defined as the mean of sensitivity and specificity, were calculated using a predicted probability threshold of 0.5. The Brier score was used to assess prediction error.

**Table 1.** Predictors included in the four logistic regression models. Tissue categories: GI, skin, brain/central nervous system (CNS) and other tissues. The expanded cohort included a multi-day indicator in every model. Delivery-parameter-only models were fitted on identical observations; comparisons assessed incremental prediction, not causality.

| Model | Dose per fraction | Tissue | DPP | Average dose rate |
|---|---|---|---|---|
| Baseline | ✓ | ✓ | — | — |
| Baseline + DPP | ✓ | ✓ | ✓ | — |
| Baseline + average rate | ✓ | ✓ | — | ✓ |
| Baseline + DPP + average rate | ✓ | ✓ | ✓ | ✓ |

Cutoffs were selected using publication-grouped five-fold cross-validation. Candidate boundaries were midpoints between adjacent values or outside the observed range, tested in both directions. Each training fold maximised balanced accuracy or sensitivity subject to specificity ≥80%; rules were evaluated on equally weighted held-out comparisons. Tissue-specific dose analyses compared a single boundary with bounded or open-ended intervals (supplementary section S13 and table S12). Full-data cutoffs were descriptive.

Except for the detectability sensitivities described in section 2.7, predictive 95% confidence intervals used 1,000 bootstrap resamples of publication/shared-cohort groups from fixed held-out predictions, excluding single-class resamples. Paired differences used identical resamples. Intervals did not include uncertainty from model refitting or cutoff reselection. Analyses were repeated under alternative NTS definitions with unchanged five-fold group assignments; full-data cutoffs were reselected after omitting each publication group (supplementary sections S5.1–S5.3).

Secondary analyses compared intrapulse rate with DPP on a common subset, restricted analyses to pulse trains, and examined schedule inclusion, quadratic log-dose terms, equal publication weights and exclusion of qualified estimates. These comparisons were specified before model fitting. Subgroup models required ≥12 independent groups, including ≥4 contributing each outcome class (supplementary section S5.3). An exploratory influence analysis omitted each publication group and refitted the baseline and baseline-plus-DPP models by the same LOPO procedure on the remaining conditions (supplementary section S18).

### 2.7. Detectability sensitivity analysis

We examined whether the predictive value of DPP was sensitive to the presence and extent of conventional radiation injury. All 185 unique irradiation conditions in the expanded matched electron cohort, including 156 single-exposure conditions, were assessed against two eligibility criteria before model fitting.

The first analysis was restricted to conditions with documented conventional normal tissue injury in at least one associated endpoint. Eligibility was based on direct measurements of injury or deterioration relative to an untreated or baseline reference. High doses, differences between FLASH and CONV, changes in molecular markers alone and injury categories assigned during data extraction were not sufficient evidence of conventional tissue injury.

The second analysis was restricted to conditions with an observed conventional adverse event probability between zero and one, excluding both extremes, for a binary endpoint defined in the original study. Exact event counts were distinguished from qualified estimates derived from survival, event probabilities and functional categories. Continuous measurements and ordinal scores were not converted into complication probabilities.

The baseline and baseline-plus-DPP models were refitted within each eligible subset using LOPO cross-validation, retaining the original outcome classifications, publication groups and training-fold preprocessing. Model fitting required at least 12 publication groups, with at least four groups contributing to each outcome class and both classes represented in every training fold. Predictions from the full-cohort models were also evaluated within the documented-injury subsets to distinguish changes in model performance associated with restricting the evaluation population from those associated with refitting.

Paired 95% confidence intervals were calculated from 2,000 bootstrap resamples of publication groups using the held-out predictions. These analyses assessed the sensitivity of predictive performance to the conditions included, recognising that documented conventional injury does not necessarily indicate sufficient statistical power to detect NTS.

Eligibility decisions and supporting source references are provided with the accompanying analysis data. Assessment was neither blinded to published outcomes nor independently duplicated. Full eligibility criteria and analytical methods are described in supplementary section S19.

## 3. Results

### 3.1. Evidence composition and endpoint agreement

Across 79 publications, 510 evaluable endpoint comparisons were grouped into 383 irradiation conditions. NTS was reported in at least one endpoint in 254 conditions (66.3%). Of these, 44 (17.3%) also contained endpoints without NTS. Requiring NTS in every endpoint reduced the positive proportion to 210/383 (54.8%). Tissue-specific frequencies and endpoint agreement are shown in figure 1. Full counts by tissue and schedule are provided in supplementary table S4.

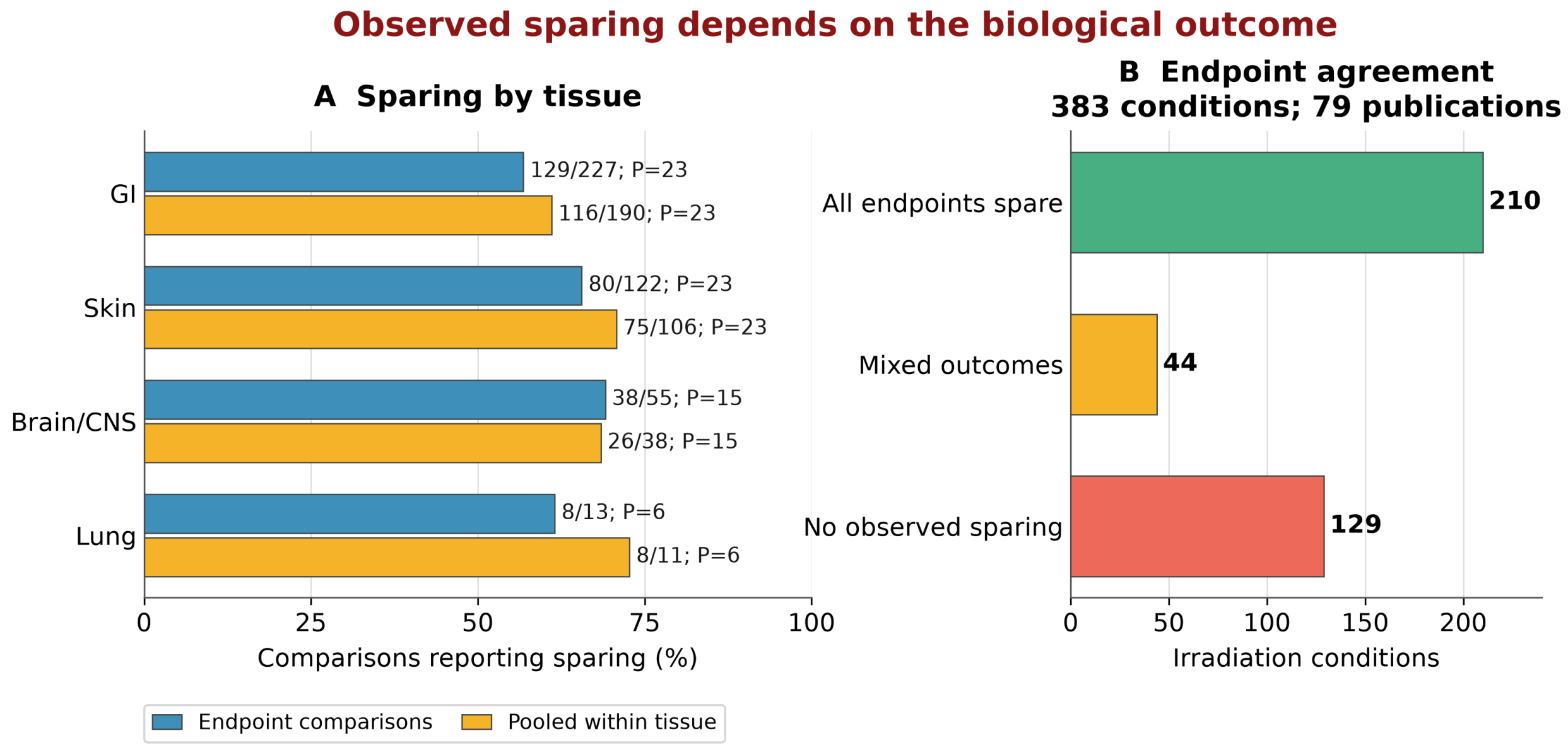

**Figure 1.** Normal tissue sparing (NTS) by tissue and endpoint agreement. (A) Endpoint and pooled condition counts by tissue (NTS/total). Tissue strata overlap. (B) Endpoint agreement in 383 conditions from 79 publications. Mixed conditions contain endpoints with and without NTS. P = publications.

### 3.2. Magnitude of normal tissue sparing

#### 3.2.1. Dose modification

Mean DMFs for single exposures, with equal weighting across publication groups, were 1.38 for skin and 1.09 for gastrointestinal (GI) injury (figure 2). The skin analysis included 11 distinct estimates from 8 publication groups, covering both acute and late injury, while the GI analysis included 7 estimates from 6 groups, all concerning acute injury. Of the 18 estimates, 15 were derived from dose–response curves or associated summaries, 1 from an extrapolated fit and 2 from approximate matched-effect comparisons. The level of injury at which DMFs were determined varied between studies. Excluding the approximate estimates reduced the mean DMFs to 1.36 for skin and 1.06 for GI injury. Further exclusion of the extrapolated fit reduced the skin DMF to 1.31, with the GI estimate remaining unchanged (supplementary section S9.3).

Paired FLASH and CONV dose series were identified in 39 of the 101 comparative publications. Of these, 38 reported tissue injury or functional outcomes, including 1 study using only hybrid FLASH/CONV irradiation, and 1 reported molecular outcomes alone. Excluding these 2 studies left 37 publications with tissue injury or functional outcomes following separate FLASH and CONV exposures. Observed responses below and above 50% in both modalities were confirmed in 12 publications, of which 7 also reported fitted dose–response curves with at least 3 dose groups per modality. Excluding the hybrid study reduced these counts to 11 and 6, respectively (supplementary section S9.3).

Sequential exclusion of individual publication groups yielded mean DMFs ranging from 1.36 to 1.41 for skin and from 1.06 to 1.11 for GI injury (supplementary table S9). No lung comparison met the isoeffect criteria. Survival estimates following pelvic irradiation are presented separately in figure S7A.

### 3.2.2. Event reduction and residual injury

FLASH had lower observed event risk in 53 of 75 matched comparisons, equal risk in 19 and higher risk in 3. Across 20 publications, 71 comparisons measured death or toxicity-related euthanasia and four measured nonfatal events. Reductions ranged from 6.7 to 100 percentage points. Residual FLASH events occurred in 40 of the 53 lower-risk comparisons, affecting 6.7–80% of evaluable animals.

Figure 3 shows all 61 GI and skin death/euthanasia comparisons by schedule. Single-exposure GI differences ranged from a 9.1-point increase to a 70-point reduction; multi-day differences ranged from a 5-point increase to a 30-point reduction. Skin reductions ranged from 0–100 points after single exposure and 0–65.7 points after multi-day treatment. The multi-day GI and skin subsets each came from one publication. Other events are shown in figure S10.

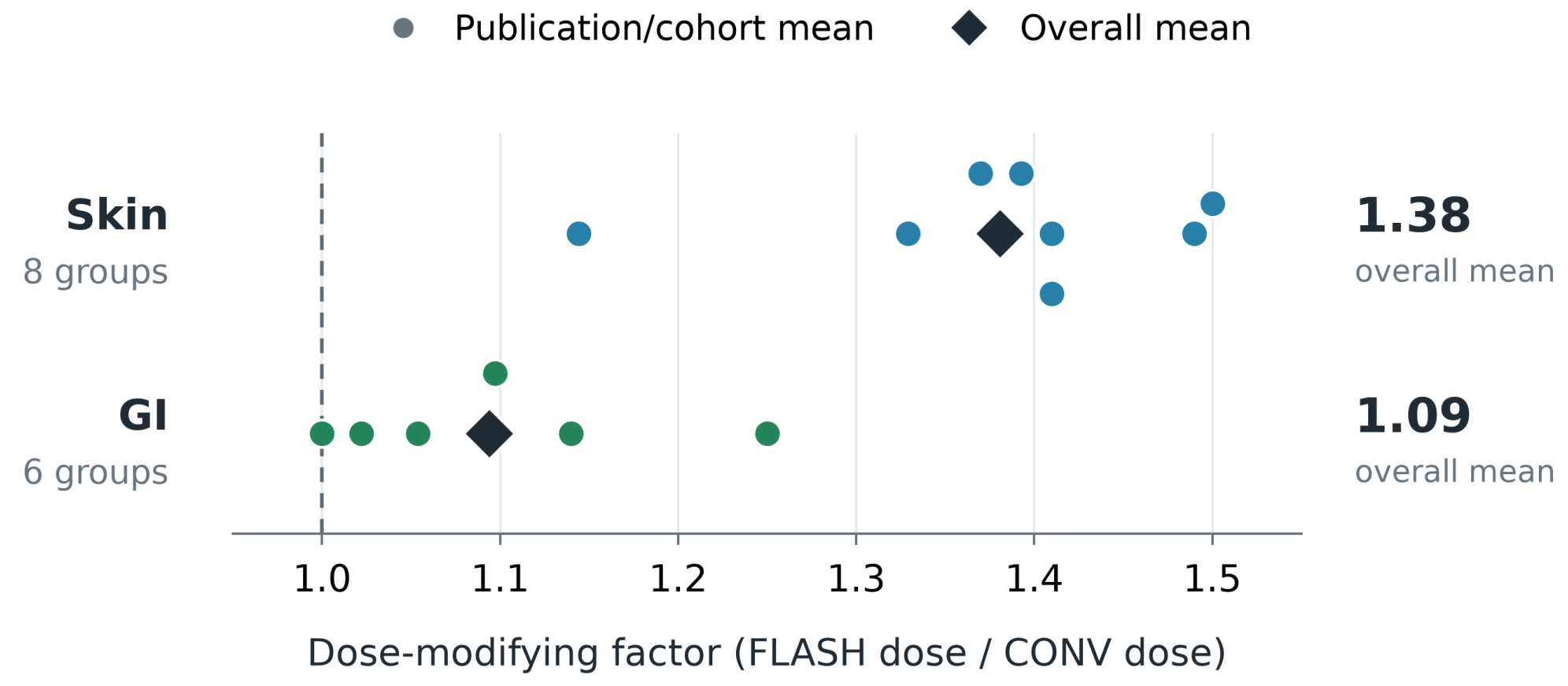

**Figure 2.** Dose modification after single exposure. Circles show publication/shared-cohort means and diamonds their equally weighted mean: 11 skin estimates from eight groups and seven GI estimates from six groups. All 18 eligible estimates are included, comprising 15 curve estimates or linked summaries, one extrapolated fit and two approximate matched-effect comparisons. DMF = FLASH/CONV dose at equivalent injury (dashed line, DMF = 1). CONV, conventional irradiation. Source estimates and provenance sensitivities are detailed in supplementary section S9 and table S9.

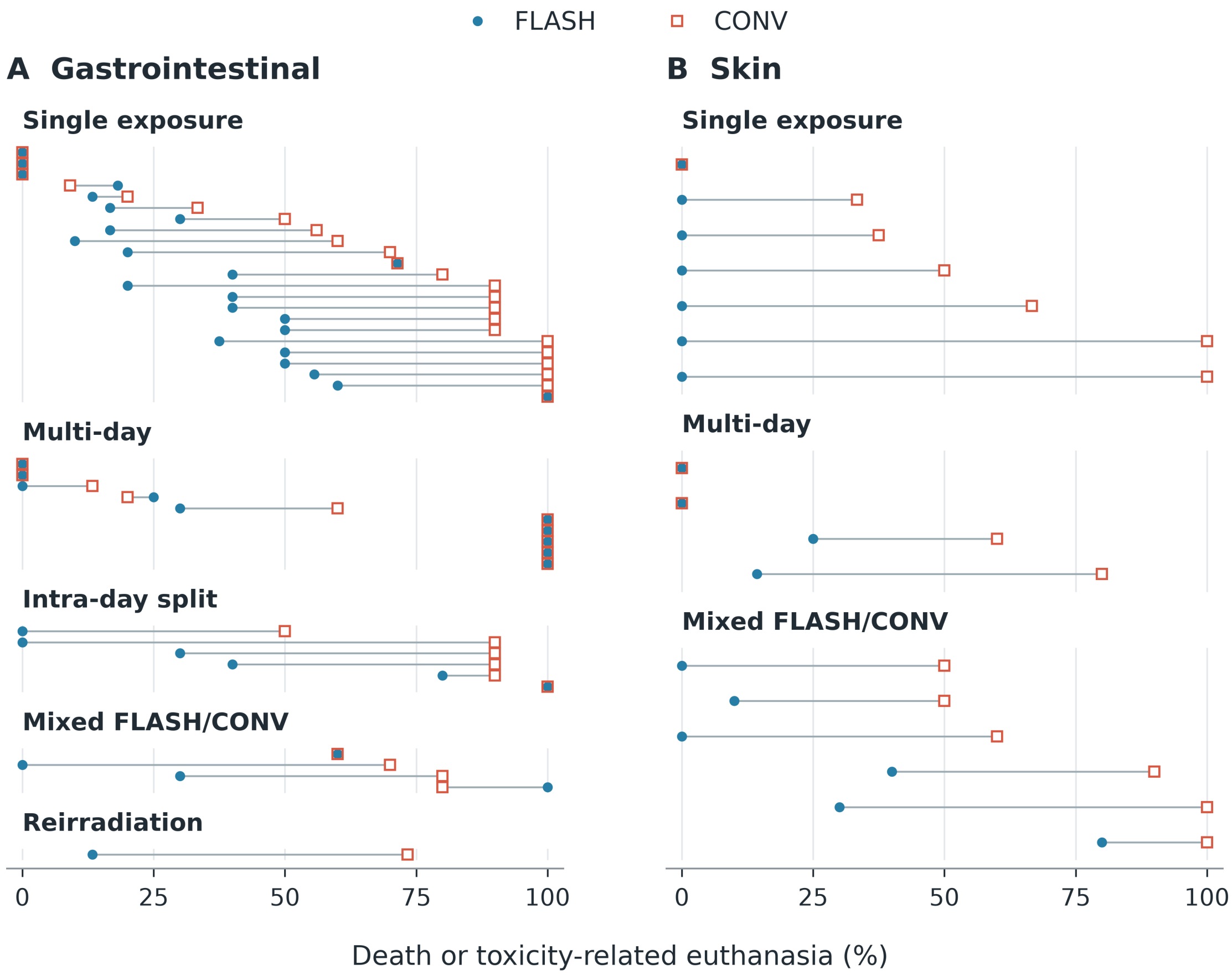

**Figure 3.** Risk of death or toxicity-related euthanasia following FLASH and CONV irradiation for GI (A) and skin (B) injury, grouped by treatment schedule. Lines connect matched comparisons, with FLASH markers to the left indicating reduced risk. Comparisons/publications from top to bottom: GI: 23/9, 10/1, 6/2, 4/1 and 1/1; skin: 7/2, 4/1 and 6/1. Follow-up times and risk-difference confidence intervals are provided in supplementary section S14.

### 3.3. Treatment schedule and fraction count

NTS occurred in 202/295 single-exposure conditions (68.5%) and 14/33 multi-day conditions (42.4%). Condition frequencies fell to 50.0% with 2–3 fractions, 40.0% with 4–6 and 16.7% with ≥7. The latter two groups comprised four and two publications, respectively (figure 4A). Intra-day splitting showed NTS in 22/30 conditions (73.3%, figure 4B).

DMFs within individual studies decreased from 1.41 after one fraction to 1.05 after eight fractions for acute skin injury, and from 1.14 after one fraction to 1.00 after ten fractions for GI injury (Kristensen et al 2025c; Sesink et al 2026a; supplementary figure S6).

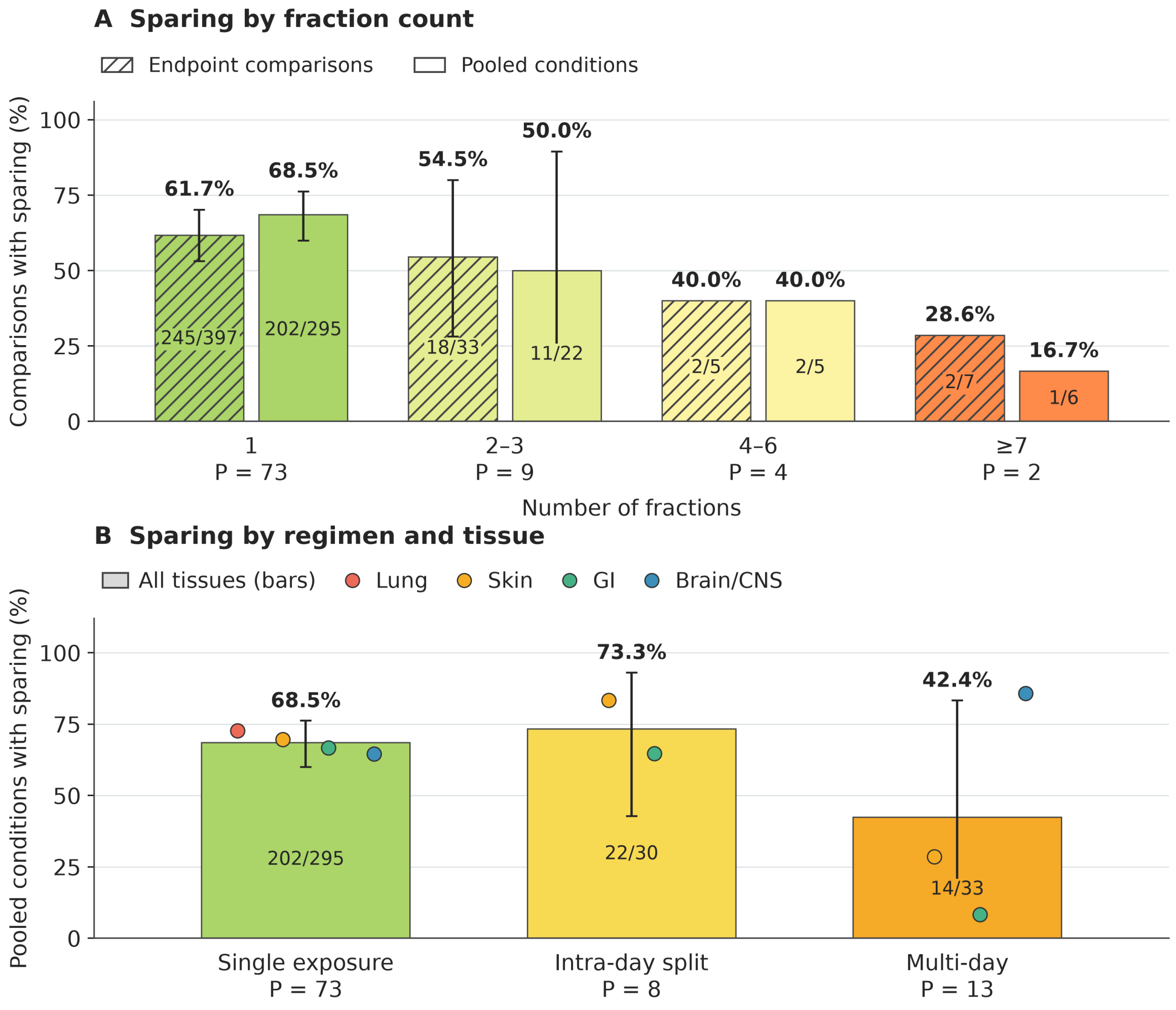

**Figure 4. NTS** by fraction count (A) and schedule/tissue (B). Hatched bars represent endpoints, solid bars pooled conditions and coloured points tissue-specific frequencies. Bar labels give NTS/total. Whiskers show 95% publication/cohort-group bootstrap intervals, omitted with fewer than five groups. P = publications.

### 3.4. Tissue-specific dose ranges

Single-exposure dose ranges overlapped between conditions with and without NTS in GI, skin, brain/CNS and lung (table 2A). Distributions by beam type and coverage across all tissues are shown in supplementary figure S11 and table S11, respectively.

Skin had the highest held-out balanced accuracy for a single dose boundary. The interval rule reduced accuracy in skin, left brain/CNS performance unchanged and increased GI accuracy from 36.7% to 42.3% (table 2B). Results including multi-day schedules and analyses prioritising specificity are reported in table S12.

**Table 2.** Single-exposure dose ranges (A) and dose-only prediction (B). Ranges are observed extrema for conditions pooled within tissue. Cutoffs maximise training BA. Displayed cutoffs use all eligible data, whereas performance uses held-out publication/cohort groups. n = conditions, P = publications. Lung did not meet validation criteria. Cutoff precision reflects dose grid midpoints (supplementary section S13).

**A Observed dose ranges after single exposure**

| Tissue | Conditions / publications | NTS n/N (%) | With NTS (Gy) | Without observed NTS (Gy) |
|---|---|---|---|---|
| GI | 144 / 21 | 96/144 (66.7%) | 7.5–21 | 5–17 |
| Skin | 79 / 22 | 55/79 (69.6%) | 17–66.88 | 15.6–45 |
| Brain/CNS | 31 / 12 | 20/31 (64.5%) | 8–30.4 | 10–25 |
| Lung | 11 / 6 | 8/11 (72.7%) | 10–30 | 14.57–17.5 |

**B Dose-only rules and performance in held-out publications**

| Tissue | n / P | One boundary (Gy) | Held-out BA % (95% CI) | Interval rule (Gy) | Held-out BA % (95% CI) |
|---|---|---|---|---|---|
| GI | 143 / 21 | ≥ 14.265 | 36.7 (30.1–49.4) | 11.15–12.4 | 42.3 (32.5–57.2) |
| Skin | 79 / 22 | ≥ 27.5 | 63.6 (49.4–75.9) | 27.5–42.15 | 51.1 (41.3–65.2) |
| Brain/CNS | 31 / 12 | ≤ 12 | 53.6 (42.4–70.8) | ≤ 12 | 53.6 (42.4–70.8) |

### 3.5. Experimental context

After accounting for tissue, beam type and treatment schedule, higher dose per fraction was associated with greater odds of NTS (OR per doubling 2.84, 95% CI 1.50–5.39). Average dose rate showed a weaker association after additional adjustment for dose per fraction (OR 1.10, 95% CI 0.98–1.24). However, neither association reached statistical significance after correction for multiple testing ($q = 0.068$ and 0.294, respectively).

NTS was reported in every assessment window, including >90 days (figure 5A). Tissue and endpoint composition differed between windows (supplementary figure S9B). NTS occurred in 120/172 conditions under ambient or medical air, 22/45 with supplemental oxygen and 2/6 under low oxygen. Inspired gas was unreported for 160 conditions (figure 5B). Field areas varied between experiments targeting the same tissue (figure 5C).

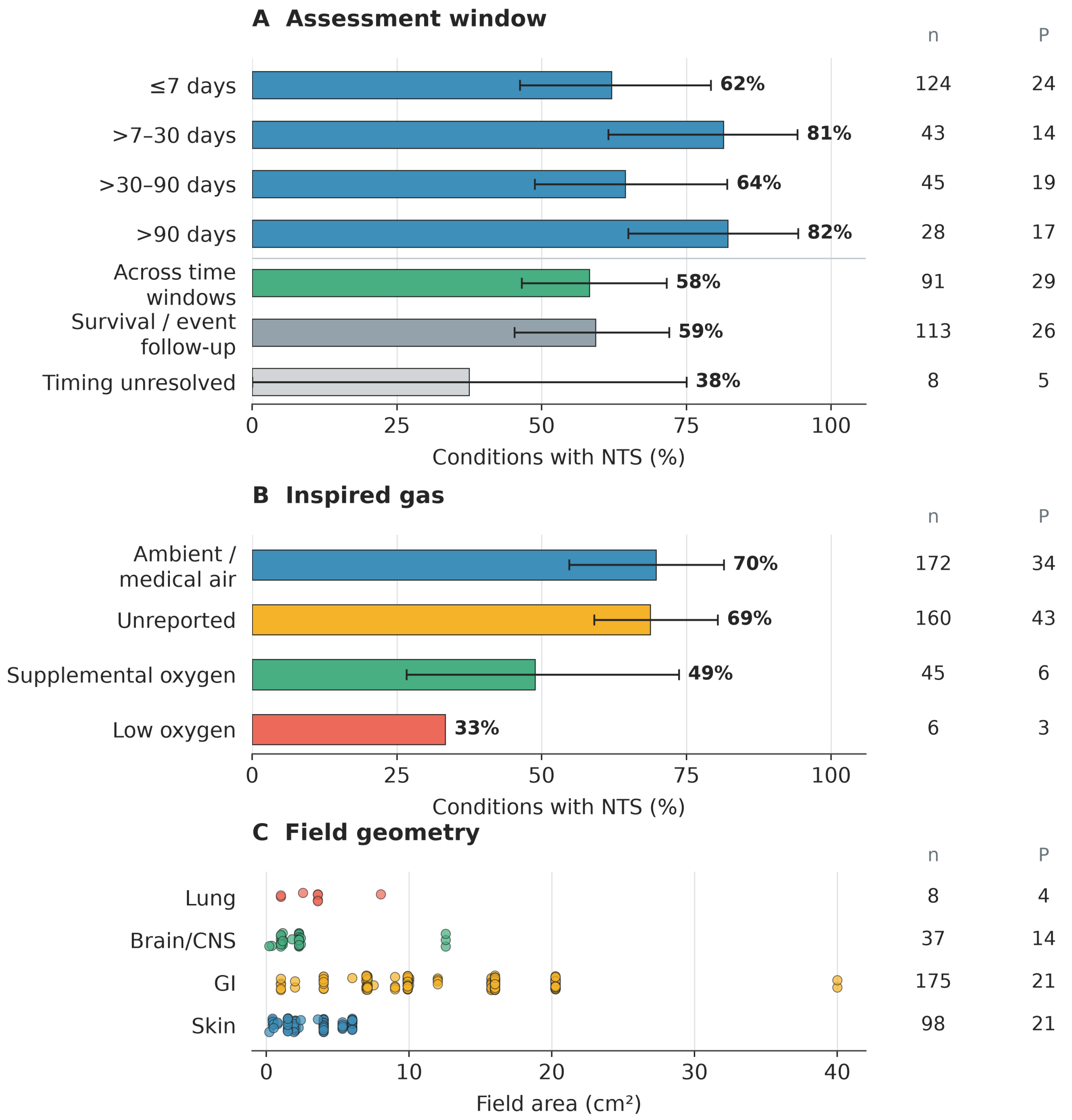

**Figure 5.** Experimental context. NTS among pooled conditions by assessment window (A) and inspired gas (B), with 95% publication-group bootstrap intervals where estimable. Conditions may contribute to several timing strata. (C) Field areas by tissue. n = conditions, P = publications. Timing definitions are detailed in supplementary section S12.

### 3.6. Machine-learning prediction across publications

Matched prediction cohorts comprised 156 single-exposure conditions from 32 publications and 185 conditions from 36 publications including homogeneous multi-day treatment. Adding DPP to the baseline increased AUROC from 0.421 to 0.583 (95% CI 0.449–0.715) for single exposures and from 0.511 to 0.683 (0.565–0.785) in the expanded cohort (figure 6, table 3A). Paired increments were 0.162 (0.008–0.271) and 0.172 (0.020–0.289), respectively.

Adding average rate alone yielded AUROCs of 0.460 and 0.581; adding both parameters yielded 0.565 and 0.635. DPP alone yielded 0.613 and 0.605 on the same conditions. Models adding DPP to the baseline had sensitivities of 92.6% and 89.8%, but specificities of 31.2% and 41.8%. Probability error was lower than for the baselines (table 3A). Models using delivery parameters alone and paired differences are reported in supplementary tables S14–S15. Parameter-available and curve-inclusive analyses are in table S5.

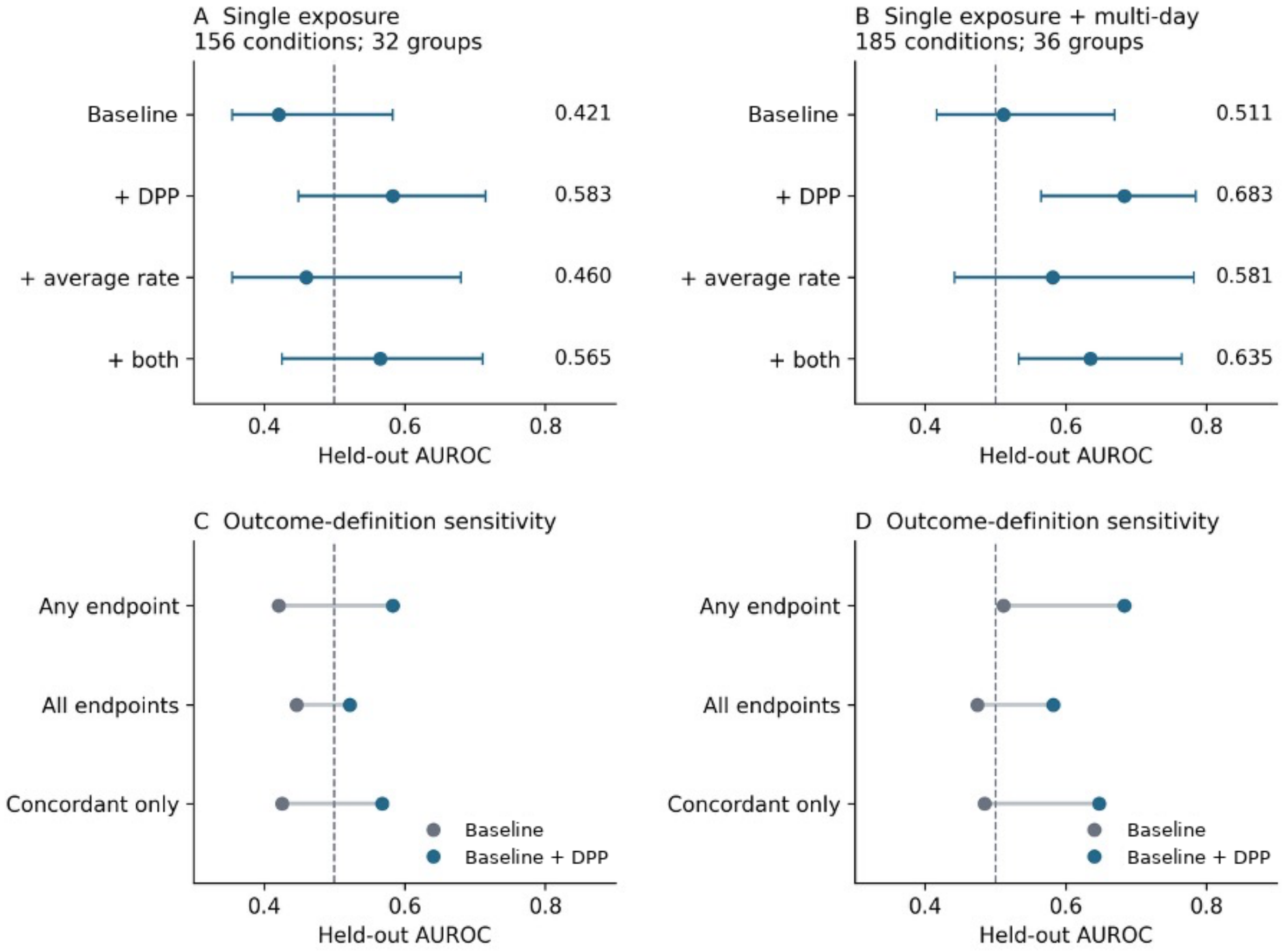

**Figure 6.** Incremental prediction and outcome-definition sensitivity. (A, B) Delivery parameters are added to dose per fraction and tissue, plus a multi-day indicator in B. Points show pooled LOPO AUROC with 95% publication-group bootstrap intervals. (C, D) Baseline and baseline-plus-DPP point estimates under three outcome definitions, from table S16. Any- and all-endpoint models use 156/32 (C) and 185/36 (D) conditions/groups; concordant-only models use 144/29 and 171/31. Comparisons within each definition share observations and validation groups. Dashed lines mark AUROC = 0.5.

When all three delivery parameters were available, intrapulse dose rate and DPP showed similar predictive performance. The respective AUROCs were 0.648 and 0.639 for single exposures (139 conditions, 25 publications) and 0.586 and 0.620 with multi-day treatment included (167 conditions, 28 publications). Confidence intervals for paired AUROC differences included zero in both cohorts (supplementary table S17).

Including quadratic dose terms, assigning equal training weights across publication groups and excluding qualified parameter estimates had little effect on baseline-plus-DPP AUROCs. Requiring NTS across all endpoints reduced AUROC to 0.522 and 0.582, compared with baseline values of 0.446 and 0.474, respectively (figure 6C,D). Restricting the analysis to conditions with concordant endpoints yielded AUROCs of 0.568 and 0.647, with corresponding baselines of 0.426 and 0.484. Restricting the analysis to pulse trains reduced baseline-plus-DPP AUROC to 0.470 and 0.615. Including a treatment schedule indicator had little effect on expanded-cohort AUROC (0.683 with versus 0.684 without). Additional sensitivity analyses are reported in supplementary section S17 and table S16.

Both models were refitted using LOPO after excluding each publication group in turn. Adding DPP improved AUROC in all 32 single-exposure analyses (range 0.067–0.235) and all 36 expanded-cohort analyses (0.093–0.201). The smallest improvements occurred following exclusion of Liu et al (2025b) and Montay-Gruel et al (2017), respectively (supplementary section S18).

**Table 3.** Electron prediction models (A) and candidate cutoffs (B). The baseline contains dose per fraction and tissue, plus a multi-day indicator in the expanded cohort. (A) Leave-one-publication-out performance at probability ≥0.5. Lower Brier scores indicate less probability error. (B) Full-data cutoffs and held-out performance of publication-grouped five-fold selection. Parentheses give 95% group-bootstrap intervals. Model cohorts require a unique dose and both delivery values. Cutoff cohorts are parameter-specific. n = comparisons, P = publications, G = validation groups.

**A. Classification and probability accuracy**

| Predictor | Brier score | Balanced accuracy (%) | Sensitivity (%) | Specificity (%) |
|---|---|---|---|---|
| **Single exposure: 156 conditions; 32 publications and validation groups** | | | | |
| Baseline | 0.237 | 49.5 | 99.1 | 0.0 |
| Baseline + DPP | 0.208 | 61.9 | 92.6 | 31.2 |
| Baseline + average rate | 0.247 | 47.2 | 86.1 | 8.3 |
| Baseline + DPP + average rate | 0.213 | 62.4 | 93.5 | 31.2 |
| **Single exposure + multi-day: 185 conditions; 36 publications and validation groups** | | | | |
| Baseline | 0.249 | 52.5 | 85.6 | 19.4 |
| Baseline + DPP | 0.205 | 65.8 | 89.8 | 41.8 |
| Baseline + average rate | 0.249 | 47.9 | 70.3 | 25.4 |
| Baseline + DPP + average rate | 0.225 | 63.2 | 83.1 | 43.3 |

**B. Threshold rules**

| Parameter | n | P | G | Full-data cutoff | Held-out balanced accuracy (%) (95% CI) |
|---|---|---|---|---|---|
| **Single exposure** | | | | | |
| DPP | 165 | 38 | 37 | ≥0.645 Gy/pulse | 58.2 (46.4–65.8) |
| Average dose rate | 174 | 35 | 35 | ≥40.25 Gy/s | 58.5 (46.1–68.8) |
| **Single exposure + multi-day** | | | | | |
| DPP | 196 | 42 | 41 | ≥3.85 Gy/pulse | 61.7 (50.5–67.9) |
| Average dose rate | 205 | 39 | 39 | ≥40.25 Gy/s | 55.4 (47.3–65.2) |

Restricting the analysis to conditions with documented conventional injury retained 114 single-exposure conditions and 140 conditions when multi-day treatments were included. Adding DPP increased AUROC from 0.540 to 0.625 and from 0.606 to 0.703, respectively. Although these improvements were smaller than those observed in the unrestricted cohorts, the resulting AUROCs were higher in both cases. The 95% confidence intervals for both improvements included zero. The 45 exclusions were primarily due to the absence of within-study healthy reference measurements. As the restriction substantially changed the representation of individual studies, these findings should be considered exploratory. Further analysis restricted to intermediate conventional response probabilities showed negligible improvement with DPP but was limited to 38 conditions across 12 publication groups (supplementary tables S19–S21).

### 3.8. Candidate thresholds and publication influence

Selecting rules for maximum BA in all eligible data gave DPP cutoffs of ≥0.645 Gy/pulse for single electron exposures and ≥3.85 Gy/pulse including multi-day schedules (figure 7, table 3B). The corresponding selection procedures achieved held-out BA of 58.2% and 61.7%. The average rate cutoff was ≥40.25 Gy/s in both cohorts, with held-out BA of 58.5% and 55.4%.

Across five training folds, DPP cutoffs ranged from 0.603 to 4.30 Gy/pulse for single exposures and 0.915 to 4.225 Gy/pulse including multi-day schedules. Average rate cutoffs ranged from 40.25–249 and 40.25–1020 Gy/s, respectively (supplementary table S7). Rules prioritising specificity and rules combining parameters are reported in supplementary table S8 and figure S4.

The DPP cutoff was unchanged after 29/37 publication-group deletions in the single-exposure cohort and 38/41 in the expanded cohort. The respective deletion ranges were 0.603–4.225 and 0.865–4.225 Gy/pulse (figure 7C,D). Omitting Grilj et al ( Grilj 2025) was the only deletion to change the average rate cutoff, from 40.25 to 49.5 Gy/s in both cohorts (supplementary figure S5).

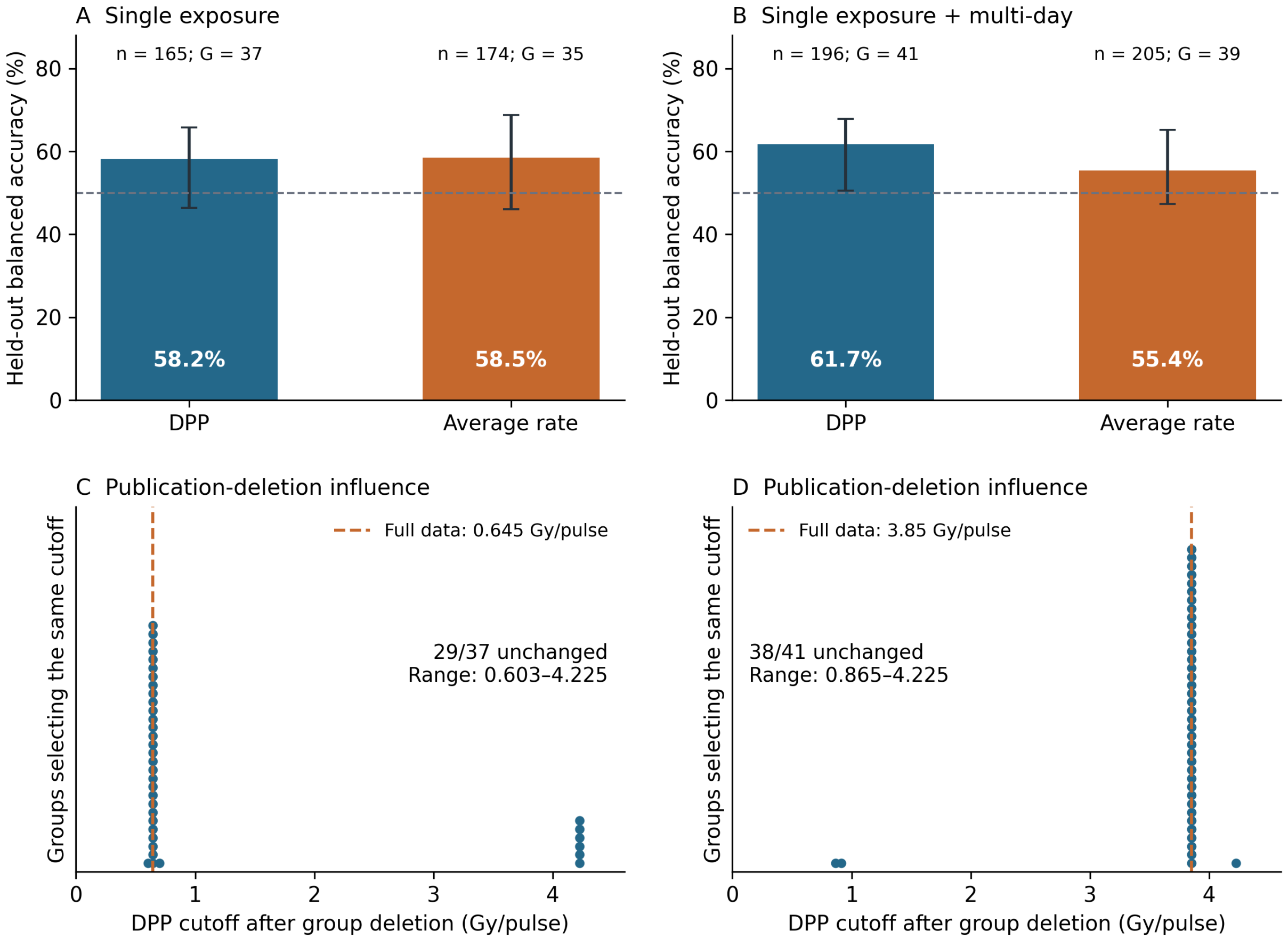

**Figure 7.** Candidate cutoff validation and publication influence. (A, B) Held-out balanced accuracy of maximum-training-BA selection under publication-grouped five-fold validation, with 95% group-bootstrap intervals; n gives comparisons and G validation groups in the parameter-specific cohorts. (C, D) DPP cutoffs after each publication-group deletion, with identical values stacked; dashed lines show full-data cutoffs of 0.645 and 3.85 Gy/pulse. Detailed average-rate influence and training-fold rules remain in supplementary figure S5 and table S7.

## 4. Discussion

This study provides a quantitative assessment of normal tissue sparing across 101 comparative in vivo FLASH publications, examining its frequency, magnitude and relationship with biological and physical parameters. NTS was reported in 66.3% of irradiation conditions, with substantial variation across tissues and treatment schedules. DPP provided additional predictive information beyond dose per fraction and tissue type, supporting the importance of temporal beam structure in FLASH radiotherapy. However, the variation in candidate delivery thresholds highlights the difficulty of establishing universal physical requirements for sparing.

The mean single-exposure DMFs of 1.38 for skin and 1.09 for GI injury indicate substantial potential for increasing the therapeutic window, particularly for skin irradiation. Even modest shifts in dose–response curves can produce considerable reductions in toxicity within the steepest region of the curve (Böhlen et al 2022). Differences in DMF between tissues may reflect biological sensitivity, irradiation geometry, endpoint selection and assessment timing. Importantly, adverse events remained common even under conditions where FLASH reduced their incidence, demonstrating that sparing frequently represents a reduction in injury rather than complete protection. This distinction is particularly relevant when considering clinical dose escalation.

Treatment schedule was an important factor associated with NTS, with sparing reported less frequently following multi-day treatment than single exposures (42.4% versus 68.5%). This is consistent with experimental evidence of reduced dose modification following fractionated skin and GI irradiation (Kristensen et al 2025c; Sesink et al 2026a). Nevertheless, protection has been demonstrated under selected fractionated protocols (Limoli et al 2023; Paillas et al 2026), and the relatively frequent sparing observed following intra-day dose splitting warrants further investigation. Understanding the combined influence of fraction size, fraction number and interfraction interval will be important for extending FLASH to conventional clinical treatment schedules.

The detection of NTS depends on the underlying dose–response relationship. At doses producing minimal or saturated injury, differences between FLASH and CONV may be difficult to detect despite an underlying shift in the dose–response curve (Böhlen et al 2022). Only 39 of the 101 publications included paired dose series, and just 12 demonstrated responses below and above 50% in both modalities. More comprehensive sampling of dose–response curves would improve estimates of sparing magnitude and help distinguish the absence of protection from limitations in its detection.

DPP provided greater incremental predictive information than average dose rate after accounting for dose per fraction and tissue type. Improvements in held-out AUROC were retained following the individual exclusion of every publication group, demonstrating that this finding was not dependent on a single study. Adding DPP also improved prediction within the documented-injury cohorts, although the incremental improvements were smaller and their confidence intervals included zero. The similar predictive performance of DPP and intrapulse dose rate is consistent with their physical relationship and highlights the need for experiments that independently vary temporal delivery parameters. These findings complement controlled studies demonstrating the influence of both DPP and average dose rate on biological response (Liu et al 2025b; Grilj et al 2025).

Recent studies have advanced the quantitative modelling of FLASH sparing through isoeffect-based temporal models (Hecker et al 2026) and endpoint-specific regression analyses (Colizzi et al 2026). The present study extends this work through broader coverage of tissues, endpoints and treatment schedules, combining quantitative measures of sparing with an assessment of delivery parameters beyond established biological predictors. Matched-cohort comparisons, publication-level validation and sensitivity analyses allowed the incremental predictive value of delivery parameters and the generalisability of candidate thresholds to be examined across heterogeneous experimental conditions.

The DPP threshold increased from 0.645 to 3.85 Gy/pulse when multi-day treatments were included, whereas the average dose rate threshold remained close to the commonly adopted value of 40 Gy/s. The limited held-out discrimination of both thresholds suggests that neither parameter alone adequately describes the conditions required for NTS. Rather than seeking a universal delivery threshold, future investigations should establish physical requirements within defined biological and treatment settings, accounting for interactions between dose, temporal beam structure and fractionation.

Physiological conditions may also contribute to variation in FLASH sparing. NTS was reported less frequently under supplemental oxygen, consistent with experimental evidence of tissue-dependent oxygen effects (Sesink et al 2026b). Anaesthesia and immobilisation may further influence tissue response (Kristensen et al 2025b; Tavakkoli et al 2025). Characterising these conditions alongside physical delivery parameters is important for reproducibility and comparison between experimental centres.

The principal limitations of this study arise from heterogeneity in experimental design, incomplete reporting and the availability of independent publication groups. The classification of NTS from reported outcomes is susceptible to publication bias, endpoint selection and differences in statistical power. Quantitative estimates of sparing were limited by the availability of paired dose–response data and differences in endpoint definition and assessment timing. The prediction analyses were further constrained by incomplete delivery parameter reporting, correlations between physical parameters and the limited number of independent experimental cohorts. Consequently, the observed associations and candidate thresholds should be interpreted as a basis for experimental investigation rather than validated delivery criteria.

Improved reporting of physical delivery parameters, physiological conditions and biological outcomes would facilitate more reliable comparisons between studies. The reporting recommendations presented in table 4 complement existing FLASH dosimetry recommendations (Böhlen et al 2024) and ARRIVE 2.0 guidelines (Percie du Sert et al 2020). Future studies should prioritise paired dose–response experiments, independent variation of temporal delivery parameters, systematic evaluation of fractionation and assessment of long-term functional outcomes alongside tumour control. Together, these developments would support the establishment of reproducible FLASH delivery criteria and the translation of preclinical sparing into clinically meaningful improvements in the therapeutic ratio.

Future studies should build on the observed dose modification and reductions in toxicity through paired FLASH and CONV dose–response experiments, particularly in skin and GI models under single-exposure and hypofractionated schedules. Independent variation of temporal delivery parameters, fraction size, fraction number and interfraction interval would help establish the conditions required to retain NTS. Consistent functional endpoints, longer follow-up and replication across centres would facilitate experimental validation of candidate delivery thresholds and dose ranges. Concurrent assessment of tumour control will be essential to determine whether normal tissue sparing translates into a clinically meaningful improvement in the therapeutic ratio.

**Table 4.** Proposed reporting parameters for comparative in vivo FLASH studies.

| Reporting domain | Information to report |
|---|---|
| Model and cohort | Species, strain, sex, age or weight, tissue, health/tumour status, independent cohorts and shared controls. |
| Study design | FLASH and conventional comparators, dose/schedule matching, allocation, blinding, exclusions and sample-size rationale. |
| Dose and geometry | Beam type and energy, delivered dose and reference location; dose distribution, field geometry and irradiated volume. |
| Dosimetry | Detectors and calibration, dose-rate corrections, independent checks and measurement uncertainty. |
| Temporal delivery | Dose per pulse, pulse width, pulse frequency and count, exposure duration, average and intrapulse dose rates; definitions of timing intervals. |
| Scanned/spill delivery | Spot/spill timing, switching gaps, local dose-rate definitions and dose–time logs. |
| Treatment schedule | Total and per-fraction dose, number of fractions, inter-fraction and within-session intervals; mixed delivery and reirradiation. |
| Physiology | Anaesthesia/analgesia, inspired oxygen and delivery method, immobilisation and interventions, oxygenation/temperature |
| Outcomes and timing | All endpoints (including unchanged/worse), scoring and NTS criteria, assessment times and follow-up. |
| Animals and events | Allocated/evaluable/excluded animals by arm and time; event counts, euthanasia criteria, censoring and survivor restrictions. |
| Effect magnitude | Matched outcomes with units, variability and uncertainty; for DMFs, ratio direction, injury level, fitted curves and time point. |

## 5. Conclusion

This study demonstrates substantial variation in the frequency and magnitude of FLASH normal tissue sparing across tissues and treatment schedules. Quantitative evidence of dose modification and reduced toxicity supports the potential for improving the therapeutic ratio, particularly in skin and GI irradiation. DPP provided greater predictive information than average dose rate beyond dose per fraction and tissue type, highlighting the importance of temporal beam structure. The dependence of sparing and candidate delivery thresholds on biological and treatment conditions suggests that universal physical criteria alone are unlikely to adequately describe the FLASH effect. These findings provide a quantitative basis for designing experiments to establish tissue- and schedule-specific delivery requirements and advance the development of reproducible, clinically relevant FLASH radiotherapy protocols.

**Acknowledgements**

The authors gratefully acknowledge the FLASH radiotherapy research community for their contributions to the experimental literature that formed the basis of this study.

**Funding statement**

This work was supported by the John and Ann Doerr Endowment in Breast Surgery and the Huaju-Ju and Launterng Wang Endowment in Breast Surgery (F.M.D.); the National Cancer Institute of the National Institutes of Health (grant 1R37CA316471-01, X.D.); the American Cancer Society Research Scholar Grant (RSG-25-1411361-01-CDP, X.D.); and the National Institutes of Health (grant R01CA266673, B.W.L.).

**AI disclosure**

OpenAI Codex was used to assist with literature data extraction, assessment of study eligibility and reported outcomes, and development of analysis code. The authors retain full responsibility for the manuscript preparation and writing, scientific content and interpretation of the article.

**Conflicts of interest**

B.W.L. has received research support and lecture honoraria from Mevion. B.W.L. is a co-founder and board member of TibaRay.

## Ethical statement

This study synthesised published aggregate data. No new experiments involving humans or animals were conducted.

## Data and code availability

The datasets and analysis code used in this study are available from the corresponding author upon reasonable request. The supplementary material provides detailed extraction criteria, outcome definitions, additional statistical methods, extended results and sensitivity analyses. An interactive summary of the FLASH evidence base is available at https://lucywhitmore96.github.io/flash-explorer/.

## References

Benjamini, Y. and Hochberg, Y., 1995. Controlling the false discovery rate: a practical and powerful approach to multiple testing. *Journal of the Royal statistical society: series B (Methodological),* 57(1), pp.289-300.

Böhlen, T.T., Germond, J.F., Bourhis, J., Vozenin, M.C., Ozsahin, E.M., Bochud, F., Bailat, C. and Moeckli, R., 2022. Normal tissue sparing by FLASH as a function of single-fraction dose: a quantitative analysis. *International Journal of Radiation Oncology* Biology* Physics*, 114(5), pp.1032-1044.

Böhlen, T.T., Psoroulas, S., Aylward, J.D., Beddar, S., Douralis, A., Delpon, G., Garibaldi, C., Gasparini, A., Schüler, E., Stephan, F. and Moeckli, R., 2024. Recording and reporting of ultra-high dose rate "FLASH" delivery for preclinical and clinical settings. *Radiotherapy and Oncology*, 200, p.110507.

Colizzi, I., Toschini, M., Lomax, A.J. and Psoroulas, S., 2026. Systematic analysis of biological endpoint variability and implications for quantitative modeling of the FLASH sparing effect. *Physics and Imaging in Radiation Oncology*, p.100915.

Esplen, N., Mendonca, M.S. and Bazalova-Carter, M., 2020. Physics and biology of ultrahigh dose-rate (FLASH) radiotherapy: a topical review. *Physics in Medicine & Biology*, 65(23), p.23TR03.

Favaudon, V., Caplier, L., Monceau, V., Pouzoulet, F., Sayarath, M., Fouillade, C., Poupon, M.F., Brito, I., Hupé, P., Bourhis, J. and Hall, J., 2014. Ultrahigh dose-rate FLASH irradiation increases the differential response between normal and tumor tissue in mice. *Science translational medicine*, 6(245), pp.245ra93-245ra93.

Grilj, P.V., Zayas, A.V., Sesink, P.A., Devanand, P.P., Repáraz, P.D., Paisley, R., Sprengers, K., Passelli, P.K., Pernot, J., Böhlen, P.T.T. and Geyer, P.R.W., 2025. Average dose rate is the major temporal beam structure parameter for preserving murine intestines with pulsed electron FLASH radiation therapy. *International Journal of Radiation Oncology* Biology* Physics*, 123(2), pp.593-601.

Hecker, Á., Fix, M.K., Manser, P., Poulsen, P.R., Moeckli, R. and Böhlen, T.T., 2026. Data-driven evaluation of quantitative metrics and models for normal-tissue sparing in FLASH radiotherapy. *Physics in Medicine and Biology.*

Ilina, A., Thomas, W.S., Cao, X., Reed, M.S., Jarvis, K., Van Der Kogel, A., Van Asselt, N., Culberson, W.S. and Pogue, B.W., 2025. FLASH effect is diminished by daily fractionation of electron RT in mouse skin. *Physics in Medicine & Biology*, 70(23), p.235020.

Kim, B.K., Kim, P.M.M., Skoufos, P.G., Diffenderfer, P.E.S., Motlagh, P.S.A.O., Kokkorakis, B.M., Koliaki, B.I., Morcos, B.G., Shoniyozov, P.K., Griffin, M.J. and Hatzigeorgiou, P.A.G., 2024. FLASH proton radiation therapy mitigates inflammatory and fibrotic pathways and preserves cardiac function in a preclinical mouse model of radiation-induced heart disease. *International Journal of Radiation Oncology* Biology* Physics*, 119(4), pp.1234-1247.

Kristensen, L., Rohrer, S., Hoffmann, L., Præstegaard, L.H., Ankjærgaard, C., Andersen, C.E., Kanouta, E., Johansen, J.G., Sahlertz, M., Nijkamp, J. and Poulsen, P.R., 2025. Electron vs proton FLASH radiation on murine skin toxicity. *Radiotherapy and Oncology*, 206, p.110796.

Kristensen, L., Overgaard, C., Johansen, J.G., Hansen, A.H., Bassler, N., Poulsen, P.R. and Sørensen, B.S., 2025. Fixation method influences FLASH skin sparing in an in vivo leg model. *Acta Oncologica*, 64, p.43972.

Kristensen, L., Rohrer, S., Johansen, J.G., Hoffmann, L., Præstegaard, L.H., Hansen, A.H., Poulsen, P.R. and Sørensen, B.S., 2025. Fractionation increasingly reduces FLASH sparing for acute murine skin damage. *Radiotherapy and Oncology*, p.111209.

Liang, K.Y. and Zeger, S.L., 1986. Longitudinal data analysis using generalized linear models. *biometrika*, pp.13-22.

Limoli, C.L. and Vozenin, M.C., 2023. Reinventing radiobiology in the light of FLASH radiotherapy. *Annual Review of Cancer Biology*, 7(1), pp.1-21.

Limoli, C.L., Kramár, E.A., Almeida, A., Petit, B., Grilj, V., Baulch, J.E., Ballesteros-Zebadua, P., Loo Jr, B.W., Wood, M.A. and Vozenin, M.C., 2023. The sparing effect of FLASH-RT on synaptic plasticity is maintained in mice with standard fractionation. *Radiotherapy and Oncology*, 186, p.109767.

Liu, K., Titt, U., Esplen, N., Connell, L., Konradsson, E., Yang, M., Wang, X., Takaoka, T., Li, Z., Koong, A.C. and Mitra, D., 2025. Discordance in acute gastrointestinal toxicity between synchrotron-based proton and linac-based electron ultra-high dose rate irradiation. *International Journal of Radiation Oncology* Biology* Physics*, 122(2), pp.491-501.

Liu, K., Waldrop, T., Aguilar, E., Mims, N., Neill, D., Delahoussaye, A., Li, Z., Swanson, D., Lin, S.H., Koong, A.C. and Taniguchi, C.M., 2025. Redefining FLASH radiation therapy: the impact of mean dose rate and dose per pulse in the gastrointestinal tract. *International Journal of Radiation Oncology* Biology* Physics,* 121(4), pp.1063-1076.

Montay-Gruel, P., Petersson, K., Jaccard, M., Boivin, G., Germond, J.F., Petit, B., Doenlen, R., Favaudon, V., Bochud, F., Bailat, C. and Bourhis, J., 2017. Irradiation in a flash: Unique sparing of memory in mice after whole brain irradiation with dose rates above 100 Gy/s. *Radiotherapy and Oncology*, 124(3), pp.365-369.

Newcombe, R.G., 1998. Interval estimation for the difference between independent proportions: comparison of eleven methods. *Statistics in medicine*, 17(8), pp.873-890.

Paillas, S., Then, C.K., Suárez-Bonnet, A., Priestnall, S.L., Kumaran, G., Ruan, J.L., Tullis, I.D., Olcina, M.M., Bottomley, M.J. and Petersson, K., 2026. Radiation dose-dependent skin toxicity and therapeutic advantage of FLASH radiotherapy via immune modulation. *International Journal of Radiation Oncology* Biology* Physics*.

Percie du Sert, N., Hurst, V., Ahluwalia, A., Alam, S., Avey, M.T., Baker, M., Browne, W.J., Clark, A., Cuthill, I.C., Dirnagl, U. and Emerson, M., 2020. The ARRIVE guidelines 2.0: Updated guidelines for reporting animal research. *PLoS biology*, 18(7), p.e3000410.

Ruan, J.L., Lee, C., Wouters, S., Tullis, I.D., Verslegers, M., Mysara, M., Then, C.K., Smart, S.C., Hill, M.A., Muschel, R.J. and Giaccia, A.J., 2021. Irradiation at ultra-high (FLASH) dose rates reduces acute normal tissue toxicity in the mouse gastrointestinal system. *International Journal of Radiation Oncology* Biology* Physics*, 111(5), pp.1250-1261.

Sesink, A., Geyer, R., Devanand, P., Böhlen, T.T., Soutter, L., Moeckli, R., Bailat, C., Herrera, F.G. and Grilj, V., 2025. Decrease in dose per fraction impairs the FLASH sparing effect in murine intestine model. *Radiotherapy and Oncology*, p.111262.

Sesink, A., Soutter, L., Geyer, R.W., Böhlen, T.T., Bailat, C. and Grilj, V., 2026. Impact of increased oxygen concentration on the FLASH sparing effect in mice is tissue dependent. *British Journal of Radiology*, 99(1178), pp.254-262.

Sesink, A., Soutter, L., Böhlen, T.T., Geyer, R., Moeckli, R., Bailat, C. and Grilj, V., 2026. Combined ultra-high and conventional dose rate irradiation spares murine normal skin and intestine from radiation induced acute toxicity. *Radiotherapy and Oncology*, p.111747.

Shi, X., Yang, Y., Zhang, W., Wang, J., Xiao, D., Ren, H., Wang, T., Gao, F., Liu, Z., Zhou, K. and Li, P., 2022. FLASH X-ray spares intestinal crypts from pyroptosis initiated by cGAS-STING activation upon radioimmunotherapy. *Proceedings of the National Academy of Sciences*, 119(43), p.e2208506119.

Simmons, D.A., Lartey, F.M., Schüler, E., Rafat, M., King, G., Kim, A., Ko, R., Semaan, S., Gonzalez, S., Jenkins, M. and Pradhan, P., 2019. Reduced cognitive deficits after FLASH irradiation of whole mouse brain are associated with less hippocampal dendritic spine loss and neuroinflammation. *Radiotherapy and Oncology*, 139, pp.4-10.

Sørensen, B.S., Sitarz, M.K., Ankjærgaard, C., Johansen, J., Andersen, C.E., Kanouta, E., Overgaard, C., Grau, C. and Poulsen, P., 2022. In vivo validation and tissue sparing factor for acute damage of pencil beam scanning proton FLASH. *Radiotherapy and Oncology*, 167, pp.109-115.

Sørensen, B.S., Sitarz, M.K., Ankjærgaard, C., Johansen, J.G., Andersen, C.E., Kanouta, E., Grau, C. and Poulsen, P., 2022. Pencil beam scanning proton FLASH maintains tumor control while normal tissue damage is reduced in a mouse model. *Radiotherapy and Oncology*, 175, pp.178-184.

Sørensen, B.S., Kanouta, E., Ankjærgaard, C., Kristensen, L., Johansen, J.G., Sitarz, M.K., Andersen, C.E., Grau, C. and Poulsen, P., 2024. Proton FLASH: impact of dose rate and split dose on acute skin toxicity in a murine model. *International Journal of Radiation Oncology* Biology* Physics*, 120(1), pp.265-275.

Tavakkoli, A.D., Daley, W.W., Hunter, D.I., Allen, B.A., Carpenter, G.C., Gladstone, D.J., Pogue, B.W. and Hoopes, P.J., 2025. Anesthesia is a potent determinant of ultra-high dose rate sparing in the murine total abdominal irradiation model. *Frontiers in Oncology*, 15, p.1666489.

Tinganelli, W., Weber, U., Puspitasari, A., Simoniello, P., Abdollahi, A., Oppermann, J., Schuy, C., Horst, F., Helm, A., Fournier, C. and Durante, M., 2022. FLASH with carbon ions: Tumor control, normal tissue sparing, and distal metastasis in a mouse osteosarcoma model. *Radiotherapy and Oncology*, 175, pp.185-190.

Velalopoulou, A., Karagounis, I.V., Cramer, G.M., Kim, M.M., Skoufos, G., Goia, D., Hagan, S., Verginadis, I.I., Shoniyozov, K., Chiango, J. and Cerullo, M., 2021. FLASH proton radiotherapy spares normal epithelial and mesenchymal tissues while preserving sarcoma response. *Cancer research*, 81(18), pp.4808-4821.

Vozenin, M.C., De Fornel, P., Petersson, K., Favaudon, V., Jaccard, M., Germond, J.F., Petit, B., Burki, M., Ferrand, G., Patin, D. and Bouchaab, H., 2019. The advantage of FLASH radiotherapy confirmed in mini-pig and cat-cancer patients. *Clinical Cancer Research*, 25(1), pp.35-42.

Wilson, J.D., Hammond, E.M., Higgins, G.S. and Petersson, K., 2020. Ultra-high dose rate (FLASH) radiotherapy: Silver bullet or fool's gold?. *Frontiers in oncology*, 9, p.496435.

Zhu, H., Liu, S., Qiu, J., Hu, A., Zhou, W., Wang, J., Gu, W., Zhu, Y., Zha, H., Xiang, R. and Li, J., 2025. Instantaneous dose rate as a crucial factor in reducing mortality and normal tissue toxicities in murine total-body irradiation: a comparative study of dose rate combinations. *Molecular Medicine*, 31(1), p.79.